\documentstyle[12pt]{article}

\newcommand \beq{\begin{eqnarray}}
\newcommand \eeq{\end{eqnarray}}
\begin{document}
\input epsf

\def\nbfepsilon{\mbox{\boldmath$\epsilon$}}
\def\nbfgrad{\mbox{\boldmath$\grad$}}
\def\bfgamma{\mbox{\boldmath$\gamma$}}
\def\bfcalA{\mbox{\boldmath${\cal A}$}}
\def\bfcalS{\mbox{\boldmath${\cal S}$}}
\def\bfp{\mbox{\boldmath$p$}}
\def\bfv{\mbox{\boldmath$v$}}
\def\bfj{\mbox{\boldmath$j$}}
\def\bfhp{\mbox{\boldmath$\hat p$}}
\def\bfei{\mbox{\boldmath$e_i$}}
\def\bfe{\mbox{\boldmath$e$}}
\def\bfej{\mbox{\boldmath$e_j$}}
\def\bfk{\mbox{\boldmath$k$}}
\def\bfq{\mbox{\boldmath$q$}}
\def\bfR{\mbox{\boldmath$R$}}
\def\bfC{\mbox{\boldmath$C$}}
\def\bfR{\mbox{\boldmath$R$}}
\def\bfX{\mbox{\boldmath$X$}}
\def\bfx{\mbox{\boldmath$x$}}
\def\bfE{\mbox{\boldmath$E$}}
\def\bfB{\mbox{\boldmath$B$}}
\def\bfy{\mbox{\boldmath$y$}}
\def\bfr{\mbox{\boldmath$r$}}
\def\rmRe{\mbox{\rm$Re$}}
\def\rmIm{\mbox{\rm$ImR$}}

\def\bfgamma{\mbox{\boldmath$\gamma$}}
\def\bfalpha{\mbox{\boldmath$\alpha$}}
\def\bfsigma{\mbox{\boldmath$\sigma$}}
\def\bfalpha{\mbox{\boldmath$\alpha$}}
\def\bfsigma{\mbox{\boldmath$\sigma$}}
\def\bfSigma{\mbox{\boldmath$\Sigma$}}
\def\bfepsilon{\mbox{\boldmath$\epsilon$}}


\def\cad{\hbox{c'est \`a dire}}
\def\Ts{\hbox{temp\'eratures}}
\def\P{\hbox{propri\'et\'e}}
\def\Ps{\hbox{propri\'et\'es}}
\def\E{\hbox{equation}}
\def\LE{\hbox{l'\'equation}}
\def\BTD{\hbox{boucles thermiques dures}}
\def\DV{\hbox{d\'eveloppement}}
\def\DNB{\hbox{d\'eveloppement en nombre de boucles}}
\def\AI{\hbox{amplitudes 1-PI}}
\def\UR{\hbox{ultrarelativiste}}
\def\btd{\hbox{boucle thermique dure}}
\def\em{\hbox{\'electromagn\'etique}}

\def\T{\hbox{temperature}}
\def\bk{\hbox{background}}
\def\P{\mbox{\psi}}
\def\BP{\mbox{\bar\Psi}}
\def\E{\hbox{equation}}
\def\Es{\hbox{equations}}
\def\QGP{\hbox{quark-gluon plasma}}
\def\HTL{\hbox{hard thermal loops}}
\def\htl{\hbox{hard thermal loop}}
\def\se{\hbox{self-energy}}
\def\pt{\hbox{polarization tensor}}
\def\pov{\hbox{point of view}}
\def\pth{\hbox{perturbation theory}}
\def\wr{\hbox{with respect to}}
\def\fn{\hbox{function}}
\def\FN{\hbox{functions}}
\def\BN{\hbox{Bloch-Nordsieck}}
\def\vp{\mbox{$\bf v\cdot p$}}
\def\vq{\mbox{$\bf v\cdot q$}}
\def\vpq{\mbox{$\bf v\cdot(p+ q)$}}
\def\tilA{\mbox{$v\cdot A$}}
\def\tilQ{\mbox{v\cdot q}}
\def\tilQ1{\mbox{$v\cdot q_1$}}
\def\tilQ2{\mbox{$v\cdot q_2$}}
\def\bfp{\mbox{\boldmath$p$}}

\hyphenation{approxima-tions}
\hyphenation{par-ti-cu-le}
\hyphenation{par-ti-cu-les}

\hyphenation{ac-com-pa-gnees}
\hyphenation{cons-tan-te}
\hyphenation{e-lec-tro-ma-gne-ti-que}
\hyphenation{e-lec-tro-ma-gne-ti-ques}
\hyphenation{im-pe-ra-tif}

\newcommand{\theo}{th\'{e}orie\,\,}
\newcommand{\mod}{mod\`ele\,\,}
\newcommand{\mods}{mod\`eles\,\,}
\newcommand{\theos}{th\'{e}ories\,\,}

\def\square{\hbox{{$\sqcup$}\llap{$\sqcap$}}}   
\def\grad{\nabla}                               
\def\del{\partial}                              

\def\frac#1#2{{#1 \over #2}}
\def\smallfrac#1#2{{\scriptstyle {#1 \over #2}}}
\def\half{\ifinner {\scriptstyle {1 \over 2}}
   \else {1 \over 2} \fi}

\def\bra#1{\langle#1\vert}              
\def\ket#1{\vert#1\rangle}              

\def\simge{\mathrel{%
   \rlap{\raise 0.511ex \hbox{$>$}}{\lower 0.511ex \hbox{$\sim$}}}}
\def\simle{\mathrel{
   \rlap{\raise 0.511ex \hbox{$<$}}{\lower 0.511ex \hbox{$\sim$}}}}


\def\parenbar#1{{\null\!                        
   \mathop#1\limits^{\hbox{\fiverm (--)}}       
   \!\null}}                                    
\def\nunubar{\parenbar{\nu}}
\def\ppbar{\parenbar{p}}


\def\buildchar#1#2#3{{\null\!                   
   \mathop#1\limits^{#2}_{#3}                   
   \!\null}}                                    
\def\overcirc#1{\buildchar{#1}{\circ}{}}


\def\slashchar#1{\setbox0=\hbox{$#1$}           
   \dimen0=\wd0                                 
   \setbox1=\hbox{/} \dimen1=\wd1               
   \ifdim\dimen0>\dimen1                        
      \rlap{\hbox to \dimen0{\hfil/\hfil}}      
      #1                                        
   \else                                        
      \rlap{\hbox to \dimen1{\hfil$#1$\hfil}}   
      /                                         
   \fi}                                         %


\def\subrightarrow#1{
  \setbox0=\hbox{
    $\displaystyle\mathop{}
    \limits_{#1}$}
  \dimen0=\wd0
  \advance \dimen0 by .5em
  \mathrel{
    \mathop{\hbox to \dimen0{\rightarrowfill}}
       \limits_{#1}}}                           

\def\real{\mathop{\rm Re}\nolimits}     
\def\imag{\mathop{\rm Im}\nolimits}     

\def\tr{\mathop{\rm tr}\nolimits}       
\def\Tr{\mathop{\rm Tr}\nolimits}       
\def\Det{\mathop{\rm Det}\nolimits}     

\def\mod{\mathop{\rm mod}\nolimits}     
\def\wrt{\mathop{\rm wrt}\nolimits}     


\def\TeV{{\rm TeV}}                     
\def\GeV{{\rm GeV}}                     
\def\MeV{{\rm MeV}}                     
\def\KeV{{\rm KeV}}                     
\def\eV{{\rm eV}}                       

\def\mb{{\rm mb}}                       
\def\mub{\hbox{$\mu$b}}                 
\def\nb{{\rm nb}}                       
\def\pb{{\rm pb}}                       

%
\def\journal#1#2#3#4{\ {#1}{\bf #2} ({#3})\  {#4}}

\def\AdvPhys{\journal{Adv.\ Phys.}}
\def\AnnPhys{\journal{Ann.\ Phys.}}
\def\EurophysLett{\journal{Europhys.\ Lett.}}
\def\JApplPhys{\journal{J.\ Appl.\ Phys.}}
\def\JMathPhys{\journal{J.\ Math.\ Phys.}}
\def\LettNuovoCimento{\journal{Lett.\ Nuovo Cimento}}
\def\Nature{\journal{Nature}}
\def\NPA{\journal{Nucl.\ Phys.\ {\bf A}}}
\def\NPB{\journal{Nucl.\ Phys.\ {\bf B}}}
\def\NuovoCimento{\journal{Nuovo Cimento}}
\def\Physica{\journal{Physica}}
\def\PLA{\journal{Phys.\ Lett.\ {\bf A}}}
\def\PLB{\journal{Phys.\ Lett.\ {\bf B}}}
\def\PR{\journal{Phys.\ Rev.}}
\def\PRC{\journal{Phys.\ Rev.\ {\bf C}}}
\def\PRD{\journal{Phys.\ Rev.\ {\bf D}}}
\def\PRB{\journal{Phys.\ Rev.\ {\bf B}}}
\def\PRL{\journal{Phys.\ Rev.\ Lett.}}
\def\PhysRept{\journal{Phys.\ Repts.}}
\def\ProcNatlAcadSci{\journal{Proc.\ Natl.\ Acad.\ Sci.}}
\def\ProcRoySoc{\journal{Proc.\ Roy.\ Soc.\ London Ser.\ A}}
\def\RevModPhys{\journal{Rev.\ Mod.\ Phys. }}
\def\Science{\journal{Science}}
\def\SovPhysJETP{\journal{Sov.\ Phys.\ JETP }}
\def\SovPhysJETPLett{\journal{Sov.\ Phys.\ JETP Lett. }}
\def\SovJNuclPhys{\journal{Sov.\ J.\ Nucl.\ Phys. }}
\def\SovPhysDoklady{\journal{Sov.\ Phys.\ Doklady}}
\def\ZPhys{\journal{Z.\ Phys. }}
\def\ZPhysA{\journal{Z.\ Phys.\ A}}
\def\ZPhysB{\journal{Z.\ Phys.\ B}}
\def\ZPhysC{\journal{Z.\ Phys.\ C}}

\begin{titlepage}
\begin{flushright} {Saclay-T96/055}
\end{flushright}
\vspace*{1.5cm}
\begin{center}
\baselineskip=13pt
{\bf  HIGH TEMPERATURE QCD\\}
\vskip0.5cm
Edmond Iancu\\
{\it Service de Physique Th\'eorique\footnote{Laboratoire de la Direction
des
Sciences de la Mati\`ere du Commissariat \`a l'Energie
Atomique}, CE-Saclay \\ 91191 Gif-sur-Yvette, France}\\
\end{center}
\vskip 2.5cm
\begin{abstract} 

I discuss finite-temperature gauge theories
as a framework to describe the quark-gluon plasma
in the regime of high temperature
where the gauge coupling is small, $g\ll 1$. I review recent progress
in the understanding of the long-range physics, with emphasis on the collective
phenomena and their consequences for the screening
of the gauge interactions. I consider some of the infrared
divergences of the perturbation theory, and discuss
the physical mechanisms which remove these divergences.

\vspace*{7.cm}

\begin{flushleft}
Invited talk given at the XXXIth Rencontres
de Moriond \\
QCD and High Energy Hadronic Interactions\\
March 23 -- 30, 1996, Les Arcs, France
\end{flushleft}

 \end{abstract}
\end{titlepage}


\newcommand{\bibit}{\it}
\newcommand{\bibbf}{\bf}
\renewenvironment{thebibliography}[1]
        {\begin{list}{\arabic{enumi}.}
        {\usecounter{enumi}\setlength{\parsep}{0pt}
\setlength{\leftmargin .75cm}{\rightmargin 0pt}
         \setlength{\itemsep}{0pt} \settowidth
        {\labelwidth}{#1.}\sloppy}}{\end{list}}

\newpage
\setcounter{equation}{0}

\section{Introduction}
It is generally accepted that, under sufficiently
high temperatures and densities, the hadronic matter 
undergoes a phase transition to a deconfined  phase, the 
quark-gluon plasma (QGP). That such a transition
exists, it is suggested by the asymptotic freedom 
 of QCD, and by the fact that,
in a plasma phase,  the confining color forces may
 be screened by many body effects, much alike as the
ordinary electric charges get screened in electromagnetic
plasmas.  This expectation is further confirmed
by  lattice calculations which predict a phase transition
at a critical temperature $T_{cr}\sim 200$ MeV, which is accessible to
 the nowadays experiences of relativistic heavy-ion collisions.

In the high temperature limit $T\gg T_{cr}$, where asymptotic
freedom allows us to expect a weak  coupling regime  $g(T)\ll 1$, 
we can study the QGP in the framework of finite-temperature
field theory and rely, at least to lowest orders,
on a perturbative expansion in powers of $g$.
The resulting description is, in many respects, complementary
to the one offered by lattice calculations, since it allows us
 to study off-equilibrium 
evolution or dynamical properties, like the ones which
may provide plasma signatures (e.g., particle production rates).
On the other hand, the comparaisons with the lattice results,
whenever possible, are useful in order to verify to which extent the
structures and the properties identified in perturbation theory
do subsist in the lower temperature ($T\simge T_{cr}$) and 
 strong coupling ($g \simge 1$) regime, which is the
regime of direct phenomenological relevance.

In what follows, I shall review briefly some
 recent progress in the field of 
high temperature gauge theories, and also mention
 some of the open problems.

\section{Collective excitations and screening}

At very high temperatures $T\gg m_f$, we can ignore
the fermion masses $m_f$ and speak about {\it ultrarelativistic
plasmas}, either abelian (e.g., a QED plasma made by electrons,
positrons and photons) or non-abelian (the quark-gluon plasma, as
described by QCD). Indeed, the particles have typical
momenta $k\sim T$, and therefore an
ultrarelativistic dispersion relation, $E(k)=k$.
Since particles can be produced or annihilated
by thermal fluctuations, the particle number density
$\rho$ is not an independent quantity,
but it is rather related to the temperature as $\rho\sim T^3$. 
Then, the typical thermal wavelength $\lambda_T=1/k\sim 1/T$
is of the same order as the mean interparticle distance
$\bar r \sim \rho^{-1/3}\sim 1/T$, and quantum effects,
like the Pauli principle, play an important role. In particular,
in thermal equilibrium, we  have to use  the quantum distribution functions,
namely $N(E)=1/({\rm e}^{\beta E}\,-\,1)$
for bosons and $n(E)=1/({\rm e}^{\beta E} \,+\,1)$
for fermions, where $\beta\equiv 1/T$. 
Thus,
in contrast to what happens for non relativistic many body systems,
 the high temperature limit of an ultrarelativistic plasma
does not correspond to a na\"{\i}ve classical limit.

The analysis of the ultrarelativistic
 plasmas in the weak coupling limit $g\ll 1$ 
(in QED, $g = e$ is the electric charge)
reveals the emergence of collective
phenomena  over a typical
space-time scale $\lambda \sim 1/gT$,
which is large with respect to both $\bar r$ and $\lambda_T$.
Correspondingly, the collective excitations carry momenta $\sim gT$, 
and are referred as ``soft'', as opposed to the ``hard'' momenta $\sim T$
of the single particle excitations.
Since $\lambda\gg \lambda_T$,
such collective phenomena show quasi-classical features and admit
a simple theoretical description \cite{us} which generalize
the kinetic theory for
ordinary non-relativistic plasmas \cite{PhysKin}.

To introduce this description, I consider the simplest case
of an ultrarelativistic QED plasma, and study the propagation
of a slowly varying electromagnetic wave $A_\mu(x)$ (with wavelength
$\lambda \sim 1/eT$) as coupled to fluctuations in the
phase-space densities of the charged particles, to be denoted
by $n_{\mp}({\bf k}, {\bf x}, t)$ for electrons (charge $-e$)
and positrons (charge $e$), respectively. The Maxwell equation
($F_{\mu\nu}\equiv \del_\mu A_\nu - \del_\nu A_\mu$)
\beq\label{maxwell}
\del_\nu F^{\nu\mu}(x)\,=\,j^\mu(x),\eeq
involves the {\it induced current}
\beq\label{jmua}
j_\mu(x)=2e\,\int\frac{{\rm d}^3k}{(2\pi)^3}\,v_\mu\,\left[ 
 n_+({\bf k}, x)- n_-({\bf k}, x)\right],
\eeq
where the factor of 2 accounts for the spin degrees of freedom,
$x^\mu =(t,{\bf x})$, $v^\mu =(1,{\bf v})$ and ${\bf v}= {\bf k}/k$
is the velocity of the ultrarelativistic fermions.
The single-particle distribution functions
 obey the Vlasov equation\cite{PhysKin}
\beq
\label{vl}
(v\cdot\del_x) n_\pm\,
\pm\, e ({\bf  E}+{\bf  v}\times {\bf  B})
\cdot\frac{\del n_\pm}{\del{\bf  k}}\,=\,0,\eeq
which together with eqs.~(\ref{maxwell}) and (\ref{jmua})
form a closed system of equations.
In the absence of the electromagnetic field,
the plasma is in thermal equilibrium, so that
$ n_\pm({\bf k}, x)\,\to\,n(k)$. For small fields, and therefore
small off-equilibrium perturbations,
we write  $n_\pm({\bf k},x)\equiv n(k) +\delta
n_\pm({\bf k},x)$, and linearize the Vlasov equation
 to get
\beq
\label{vllin}
(v\cdot\del_x) \delta n_\pm({\bf k},x)\,=\,
\mp\, e {\bf  v}\cdot {\bf  E}(x)\frac{{\rm d} n}{{\rm d} k}.\eeq
The contribution of the  magnetic
field dropped out in the right hand side because of the isotropy of the
equilibrium state. Eq.~(\ref{vllin}) can be easily
integrated with, e.g., retarded boundary conditions,
and the resulting current may be written in momentum space as
$j^\mu(q)\,=\,\Pi^{\mu\nu}(q) A_\nu(q)$, with the {\it polarisation
tensor}
\beq\label{Pi}
\Pi_{\mu\nu}(q_0, {\bf  q})= m_{D}^2\,
\left \{-\delta_{\mu 0}\delta_{\nu 0}\,+\,q_0 \int\frac{{\rm d}\Omega}{4\pi}
\frac{v_\mu\, v_\nu} {q_0 - {\bf  v}\cdot {\bf  q}
+i\eta}\right\},\eeq
where $m_D^2= e^2 T^2/3$
and the small imaginary part
in the denominator, $i\eta$ with $\eta\to 0_+$,
reflects the retarded boundary conditions.
The angular integral $\int {\rm d}
\Omega$ runs over all the orientations of the 
unit vector  ${\bf  v}$. 

Note that in the above, seemingly {\it classical}, description of the polarization
phenomena, quantum effects entered explicitly, via the Fermi-Dirac occupation
factor $n(k)$. To reassure the reader about this apparently hybrid description,
let me emphasize that eqs.~(\ref{maxwell})--(\ref{vl})
 can be rigorously derived from  quantum field theory. They represent
 the leading order in a systematic expansion in powers of $e$
of the Dyson-Schwinger equations for thermal Green's functions
\cite{us}. In this expansion, the electric charge controls
not only the strength of the interactions, but also
the soft gradients, since $\del_x A_\mu\sim eT A_\mu$,
and similarly $\del_x n({\bf k}, x) \sim eT \,n({\bf k}, x)$.
Thus the long-wavelength, collective degrees of freedom may be treated
as classical, in contrast to the single-particle, hard degrees
of freedom, which are always quantum. Genuine quantum effects,
such as pair production, only enter the kinetic theory
 at the next-to-leading order in $e$,
on the same footing as the collision terms in the right hand
side of the Vlasov equation (\ref{vl}).


Being transverse,  $q^\mu \Pi_{\mu\nu}(q)=0$,
 the polarization tensor (\ref{Pi}) is determined by only two
independent scalar functions,
which we choose as the electric ($\Pi_l$)
and the magnetic ($\Pi_t$) components, respectively:  
\beq\label{plt}
\Pi_l(q_0, q)\equiv -\Pi_{00}(q_0, q),\qquad\,\,\Pi_t(q_0,q)\equiv
\frac{1}{2}\,(\delta^{ij}-\hat q^i \hat q^j)\Pi_{ij}(q_0,{\bf q})\,.\eeq
This choice is natural since
the medium effects distinguish between the electric (or longitudinal)
and the magnetic (or transverse) sectors of the gauge
interactions: indeed, the thermal bath
involves electric charges, but not magnetic monopoles.

This distinction is especially important when we consider the 
{\it screening effects}.
The most familiar such effect is the Debye screening
of the Coulomb interaction:  the potential between
two static pointlike sources $q_1$ and $q_2$
separated by ${\bf r}$ reads
\beq\label{Vcoul}
V(r) = q_1 q_2 \int {{\rm d}^3 q \over (2 \pi)^3 } {e^{i 
{\bf q\cdot r}}
\over q^2 + \Pi_l(0,q) } \,.\eeq
To leading order
in $e$, eqs.~(\ref{Pi}) and (\ref{plt}) yield $\Pi_l(0,q)=m_{D}^2$,
and eq.~(\ref{Vcoul}) exhibits exponential attenuation
over a typical scale $\lambda_D=1/m_D \sim 1/eT$:
$V(r)\sim {\rm e}^{-m_D r}/r$.
The quantity $m_D$ is therefore known as the ``Debye mass''.

The magnetic interactions, on the other hand, are not screened
in the static limit $q_0 \to 0\,$: $\Pi_t(0,q)=0$.
For small, but non-vanishing, frequencies,
\beq\label{trans}
\Pi_t(q_0\ll q) \simeq \,-i\,\frac{\pi}{4}\,\frac{q_0}{q}\,m_D^2\,
\eeq 
is purely imaginary, and describes the attenuation
of a time-dependent magnetic field via energy transfer
toward the charged particles (``dynamical screening''). Microscopically,
this corresponds to the absorbtion of the space-like
 photons ($q_0^2<q^2$) by the hard thermal fermions
(Landau damping) \cite{PhysKin}.

Before further discussing the consequences
of the screening effects, let me just mention that a completely
similar picture   holds in QCD as well, to leading order in $g$:
the collective {\it color} oscillations of the hard
thermal quarks and gluons are described by generalized
Vlasov-type kinetic equations \cite{us} which yield
a polarisation tensor of the form $\Pi_{\mu\nu}^{ab}(q)
=\delta^{ab}\Pi_{\mu\nu}(q)$, where $a$ and $b$ are color indices
for the adjoint representation, and $\Pi_{\mu\nu}$
is given again by eq.~(\ref{Pi}), but with a Debye mass
$ m_D^2\,=\, g^2 T^2 (N_f+2N)/6$
for $N$ colors and $N_f$ number of flavors.
Moreover, in QCD, the non-abelian gauge symmetry constrains
the induced color current $j_\mu^a(x)$ to be non-linear in the soft color 
fields $A_\mu^a(x)$, so that
we have non-trivial thermal corrections for the multi-gluon {\it vertex}
functions as well:
\beq\label{exp}
j^{a}_\mu \,=\,\Pi^{ab}_{\mu\nu}A_b^\nu
+\frac{1}{2}\, \Gamma_{\mu\nu\rho}^{abc} A_b^\nu A_c^\rho+\,...
\eeq
in symbolic notations. Finally, in ultrarelativistic plasmas,
 the bosonic and fermionic degrees of freedom play symmetrical
roles, so that we also encounter collective excitations
with {\it fermionic} quantum numbers, which can still be described
by simple kinetic equations \cite{us}.
The thermal corrections which describe the collective
behaviour at the scale $gT$  ---  like
the polarisation tensor (\ref{Pi}) and the vertex corrections
in eq.~(\ref{exp}) --- are generally dubbed ``hard thermal loops''.
This reflects the fact that, in their original derivation,
which is based on Feynman graphs for thermal QCD, they
all arise from one-loop diagrams where the external line
carry soft momenta $\sim gT$, while the internal loop
momentum is hard, $\sim T$
\cite{Klimov81,BP90,FT90}.

\section{The lifetime of the quasiparticles}

 Since the screening effects reduce the range of the gauge interactions,
their resummation greatly improve the infrared (IR) behaviour
of the perturbative expansion.  
To be more specific, let me consider the computation of the lifetimes
of the plasma excitations (either hard, or soft). 
Information about the lifetime can be obtained from the retarded propagator
$S_R(t,{\bf p})$. In many cases, this decays {\it exponentially} in time,
 $S_R(t,{\bf p})\,\sim\,{\rm e}^{-i E(p)t} {\rm e}^{ -\gamma({p}) t}$,
with a {\it damping rate} $\gamma(p)$ which is essentially the total interaction
rates of the excitation.
The quasiparticle picture is consistent as long as $\gamma \ll E$.
Let me compute $\gamma$ for a fermion with
momentum $p\sim T$ which scatters off the thermal particles
(quarks and gluons). In the Born approximation (one gluon exchange),
the interaction rate is simply $\gamma = \sigma \rho$,
where $\rho\sim T^3$ is the density of the scatterers, and $\sigma =\int
{\rm d^2}q ({\rm d}\sigma/{\rm d}q^2)$, with $q$ denoting the momentum
of the exchanged (virtual) gluon. For a bare gluon, the Rutherford
formula yields ${\rm d}\sigma/{\rm d}q^2 \sim g^4/q^4$, so that
$\gamma \sim g^4 T^3 \int ({\rm d}q /q^3)$ is quadratically
infrared divergent. Actually, the screening effects soften the
 IR behaviour, and distinguish between electric and
magnetic scattering: $\gamma=\gamma_l + \gamma_t$.
 In the electric sector, we have Debye screening,
i.e. $1/q^2 \to 1/(q^2 + m_D^2)$, and therefore a dynamical
IR cut-off $m_D \sim gT\,$: $\gamma_l \sim g^4 (T^3/m_D^2)\sim g^2 T$.
In the magnetic sector, on the other hand, the dynamical screening
does not completely remove the divergence, which is just
reduced to a logarithmic one:
\beq\label{G2LR}
\gamma_t &\sim& {g^4 T^3}\,
\int_{0}^{\infty}{\rm d}q  \int_{-q}^q{\rm d}q_0\,|D_t(q_0,q)|^2\nonumber\\
 &\sim& {g^4 T^3}\,\int_{0}^{\infty}{\rm d}q  \int_{-q}^q{\rm d}q_0
\,\frac{1}{q^4 + (\pi m_D^2 q_0/4q)^2} \,\sim\,
{g^2T}\int_{0}^{m_D}\frac{{\rm d}q}{q}\,.\eeq
In this equation, $D_t(q_0,q)=1/(q_0^2- q^2 - \Pi_t(q_0,q))$
is the propagator of the magnetic photon, and in writing
the second line we used eq.~(\ref{trans}) and
 retained only the leading, IR divergent, contribution to $\gamma_t$.
With an IR cut-off $\mu$, $\gamma_t \sim g^2 T \ln (m_D/\mu)$.
The remaining logarithmic divergence is due to collisions involving the
exchange of very soft, {\it quasistatic} ($q_0\to 0$) magnetic photons,
which are not screened by plasma effects.
To see that, note that the IR contribution to
 $\gamma_t$ comes from momenta $q\ll gT$,
where $|D_t(q_0,q)|^2$ is almost a delta function of $q_0$:
\beq \label{singDT}
|D_t(q_0,q)|^2\,\simeq\,
\frac{1} {q^4 + (\pi m_D^2 q_0/4q)^2}\,
\longrightarrow_{q\to 0}\,\frac{4}{ q m_D^2}\,\delta(q_0)\,.\eeq
In QCD, one generally expects the dynamical generation of a magnetic
screening mass $\sim g^2T$, by some non-perturbative mechanism.
This is supported by lattice computations, and shows up 
through infrared divergences in perturbation theory.
Then, the QCD damping rate is IR finite and $\sim g^2 T\ln (1/g)$ \cite{BP90}.
In QED, on the other hand, it is known that no magnetic screening can occur,
so that the solution of the problem must lie somewhere else.

Let me concentrate on the abelian problem from now on.
An analysis of the higher order corrections to
eq.~(\ref{G2LR}) reveals severe (power-like) IR divergences which signal
the breakdown of the perturbation theory \cite{prl}.
Because of the specific IR behaviour of the magnetic
photon propagator, eq.~(\ref{singDT}),
the leading divergences come
from multiple collisions where all the exchanged photons
are magnetic and quasistatic. They can be studied
in the framework of an effective three-dimensional theory,
which considers the interactions of the fermion
with only {\it static} ($q_0=0$)  photons with propagator
$D_t(0,{\bf q})=1/q^2$. By using the Bloch-Nordsieck approximation,
 it is possible to resum the leading IR divergences,
 and get the correct large-time ($t\gg 1/gT$) behaviour
of the fermion propagator  $S_R(t)$ \cite{prl}. This is free of IR problems
and, rather surprisingly, it  shows a {\it non-exponential} decay in time:
\beq S_R(t)\,\sim\,
\exp\{-\alpha T t \ln (m_D t)\},\eeq
 where $\alpha=e^2/4\pi$.
Since at large times $S_R(t)$ is decreasing faster than any exponential,
it follows that  the  Fourier transform 
\beq\label{SRE}
S_R(\omega)\,=\,
\int_{-\infty}^{\infty} {\rm d}t \,{\rm e}^{-i\omega t}
S_R(t)\,\eeq  exists 
for {\it any} complex energy $\omega$. Thus,  the retarded propagator
 $S_R(\omega)$ has no singularity at the mass-shell.
  The associated spectral density $\rho(\omega)\propto
 {\rm Im}\, S_R(\omega)$ retains the shape
of a {\it resonance}  strongly peaked around the perturbative mass-shell 
$\omega \sim E(p)$, 
with a typical width of order $\sim g^2T \ln(1/g)$  \cite{prl}.
Thus, the quasiparticles are well-defined,
even if they do not correspond to the usual,
exponential, time decay of the propagator.

\section{Conclusions}

The removal of the infrared divergences by physical mechanisms is an
important self-consistency check for high temperature gauge theories.
The computation of the damping rate illustrates both
the power and the limits of the screening effects in this sense.
They sensibly improve the infrared behaviour of the perturbation
theory, and completely remove the IR problems from the electric
sector. Still, IR divergences persist in the magnetic sector,
due to the unscreened static magnetic gluons or photons.
It has been pointed out by Baym et al. \cite{Baym90} that the
dynamical screening of the time-dependent magnetic fields,
as illustrated by eq.~(\ref{trans}), is sufficient to yield
IR finite results for many quantities of physical interest, like
transport coefficients or the collisional energy loss.
This suggests that it may be possible to further develop the kinetic
approach discussed previously in order to include collision
terms and off-shell effects, thus leading to a consistent
transport theory for the high temperature QCD plasma.

More generally, the resummation of the screening effects
in the ``hard thermal loop'' approximation
enables us with a consistent perturbative description
 of the physics at short ($\sim 1/T$) and
intermediate ($\sim 1/gT$) scales. The resulting physical
picture turns out to be quite similar for abelian or
non-abelian plasmas, but important differences occur
when going to even larger scales, $\simge 1/g^2T$.
Lattice simulations of hot QCD reveal traces of the confinement
in the long-range correlations, and these may be
associated  with the infrared divergences
encountered in perturbation theory.
In QCD, one expects these divergences to be cured by
new, non-perturbative, screening effects, which should manifest
in the magnetostatic sector at momenta $\sim g^2T$  \cite{Braaten94}.
In abelian theories, where  there is no magnetic screening,
the divergences are removed --- as we have seen on the example
of the fermion lifetime --- by further resummations
of soft photon processes to all orders in perturbation theory \cite{prl}.
Another IR problem, which is currently under investigation,
is the appearance of collinear divergences, e.g.,
in the computation of the plasma production rate for soft real photons \cite{Baier94}.
This problem is currently under investigation \cite{Rebhan95}.

Finally, one may wonder about the relevance of  perturbative QCD
for the phenomenology of heavy ion reactions.
 We have indeed evidence that in the temperature regime that is
presumably accessible to these collisions, the coupling strength is
not small, rather $g\sim 2-3$.
  Is this to say then that all the
physics described here is irrelevant? I do not believe so. It is physically
plausible, and partially supported by lattice calculations,
 that some of the structures identified at scale $gT$,
as the screening effects, may be sufficiently
robust to survive even in  a regime of parameters where the
approximations made to derive them cannot be justified.

\vskip.5cm

\noindent{\large {\bf References}}

\end{document}